\documentstyle[aps,pre,preprint]{revtex}
\begin{document}
\bibliographystyle{unsrt}

\title{
Influence of  magnetic impurities on  the  
heat capacity of nuclear spins.
}
\author{A. M. Dyugaev$^{1,2}$, Yu. N. Ovchinnikov$^{1,2}$ and P. Fulde$^1$}
\address{
$^1$ Max-Planck-Institute for Physics of Complex Systems, \\ 
Bayreuther Str. 40 Haus 16  D-01187 Dresden, Germany \\
$^2$ L. D. Landau Institute for Theoretical Physics, \\
Academy of Sciences of Russia,
Kosygin Str. 2, Moscow, 117940 Russia
}

\maketitle
\begin{abstract}

It is found  that in a wide range of temperatures and magnetic fields
even a small concentration of magnetic impurities in a sample  leads 
to a   $T^{-1}$ temperature
dependence of the nuclear heat capacity. This effect is related to  a nuclear-spin
polarization by the magnetic impurities. The parameter that controls the theory
turns out not to be the impurity concentration $C_{imp}$ but instead the  
quantity   $c_{imp} \mu_e / \mu_n$, 
where $\mu_e$ and $\mu_n$ are the magnetic moments of an electron and
a nucleus, respectively. The ratio of $\mu_e$ and $\mu_n$ is of  order of $10^3$.

\end{abstract}


 During the last years nuclear spin ordering has been observed in a
considerable number of solids. For a review see Ref. \cite{oja}. This is due to an impressive progress
in cooling nuclear spin systems and temperatures as low as  $T \sim 10^{-9} K$
have been attained. The ordering temperatures of the nuclear spin systems
are as small as $58 nK$ for $Cu$ and  $0.56 nK$ for $Ag$
 \cite{huiku},\cite{hakonen}. The 
Curie
temperature for  $AuIn_2$ is $35 \mu K$. In this system 
an interplay between nuclear magnetism and superconductivity 
has been observed \cite{reehmann}.

At such low temperatures all degrees of freedom of the solid are frozen with the
exception of  the nuclear spins.
The temperature dependence of the resistivity  is
therefore due to conduction electron-nuclear spin interactions
 \cite{dyugaev}. 
It Ref. \cite{herrmann} it was demonstrated that the nuclear-spin
susceptibility depends on the impurity concentration and that the heat
capacity in low external magnetic fields does not obey a Schottky law.
Instead, it is more close to a $1/T$ behaviour. This seems to hold for a
number of compounds  \cite{herrmann}.

The aim of this paper is to show that magnetic impurities can give the main
contribution to the heat capacity at low temperatures even if their 
concentration is very low. In the following we want to give a simple
physical argument to justify that statement  before we present a more quantitative
theory.

Consider for simplicity a system in which  the magnetic interaction between
nuclear spins as well as between an impurity and the nuclear spins is of the
dipolar form

\begin{equation}
\label{1}
V_{1,2} = \frac {{\bf \mu}{_1} {\bf \mu_2} r^2 - 3 ({\bf \mu}{_1} {\bf r} ) ( {\bf \mu}{_2} {\bf r})}{r^5}.
\end{equation}

Here ${\bf \mu}{_{1,2}}=  \mu_{1,2} {\bf S}{_{1,2}}$, ${\bf \mu}{_{1,2}}$ and ${\bf S}{_{1,2}}$ are 
 the magnetic moment and spin operators  of two particles $1, 2$  separated by a distance 
 ${\bf r}$.
The temperature  of the 
nuclear-spin ordering is accordingly  of  order $T_{cn} \simeq \mu_n^2/a^3$, where $a$ is the distance between
neighbouring nuclei. Note that  the density of sites, i.e., nuclei, is $n_n \simeq a^{-3}$.

The crucial point is that the interaction between the impurity spin and a nuclear spin is much 
larger than the one between nuclear spins since the magnetic moment $\mu_{imp}$ is much larger
than the one of the nuclei $\mu_n$, i.e., $\mu_{imp} / \mu_n \simeq 10^3$. Therefore, around each
impurity there is a volume of size  $a^3 \mu_{imp} / \mu_n$ in which the impurity-nuclei interaction 
exceeds the one among the nuclei. Consequently, if the impurity concentration $c_{imp}$ exceeds 
$\mu_n / \mu_{imp}$ the different regions with dominating impurity-nuclei interaction overlap and
influence each other. We shall consider here the clean limit, i.e., concentrations $c_{imp} < \mu_n /\mu_{imp}$.
In that case it suffices to consider a single impurity. The calculated contribution to the heat capacity
has then to be multiplied merely by $c_{imp}$. Furthermore, we shall assume that the impurity spin is kept
fixed by an applied magnetic field $H_0$, i.e., $\mu_{imp} H_0 \gg T_e$, where $T_e$ denotes
the temperature of the electron system. For $T_e \simeq 10^{-4} K$ this requires a magnetic
field of order $1$ Gauss. In $Au In_2 $  \cite{reehmann} the electron-nuclei interaction is
sufficiently strong so that the nuclear temperature $T$ and the electron temperature $T_e$ 
coincide. The effective field ${\bf H} $ acting on a nucleus  consists then of the external field
${\bf H}{_0}$ and the field set up by the impurity, i.e., 
\begin{eqnarray}
\label{2}
{\bf H} ={ \bf H}_0+{\bf H} _1\\
{\bf H}{_1}({\bf r}{_n}) = \frac { 3 ({\bf \mu}{_{imp}} {\bf r} ) {\bf r} /r^2 - {\bf \mu}{_{imp}}}{r^3}. 
\end{eqnarray}

Here  ${\bf r} = {\bf r}{_n} - {\bf r}{_{imp}}$ is the distance between nucleus and impurity. The interaction
Hamiltonian is $H_{int} = - ({\bf \mu}{_n} {\bf H})$. The partition function $Z_n$ of a nuclear spin is

\begin{equation}
Z_n = \frac { sinh (\chi( 2 S +1)) }{sinh(\chi)}, \: \chi= \frac{\mu_n H(r) }{2 T},
\end{equation}
where $S$ is the spin of the nucleus.

The specific heat contribution follows from $C_n = -T \frac { \partial^2 F_n }{\partial T^2} $
where $F_n = -T \ln Z_n $. Here we have set Boltzmann's constant $k_B=1$. This gives

\begin{equation}
\label{3}
C_n = \chi^2 ( \frac {1}{sinh^2 \chi} - \frac{(2 S +1)^2}{sinh^2 ( \chi ( 2 S + 1 ))}). 
\end{equation}

The average value of the specific heat $\bar{C_n}$ is a sum
over ${\bf r}{_n}$ multiplied by the impurity concentration. Furthermore, it
is advantageous to subtract the specific heat  $C_n^{(0)}$ of the pure material 

\begin{equation}
\label{5}
\bar{C_n} -C_n^{(0)} = n_{imp} \sum_{{\bf r}_n} (C_n (H({\bf r}_n))- C_n^{(0}(H_0) )
\end{equation}

When $ C_n^{(0)}$ is expanded in powers of  $\mu_n H_0/T$ we obtain to leading order in the 
external field
\begin{equation}
\label{6}
C_n^0 = n_n (\frac{ \mu_n H_0}{T})^2 \frac {S (S+1) }{3}  
\end{equation}
Where $n_n$ is the concentration of nuclei.

In order that a high-temperature expansion of this type does also hold for the  
nuclei close to the impurity the condition $ T \gg \mu_{imp} \mu_n /a^3 $ must be 
fulfilled. The square of the effective field is 
\begin{equation}
\label{7}
H^2({\bf r}) =  H_0^2 + \frac { 3 ({\bf \mu}{_{imp}} {\bf r} )^2 - \mu_{imp}^2 r^2}{ r^8} + \frac {2}{ r^5} 
( 3 ({\bf r}{\bf H}_0) ( {\bf \mu}{_{imp}} {\bf r} ) - ({\bf \mu}{_{imp}} {\bf H}_0) r^2)
\end{equation}

From Eqs. (\ref{5}-\ref{7}) the following expression is obtained for the specific heat
in the high-temperature regime,
\begin{equation}
\label{8}
\bar{C_n} -C_n^{(0)} = n_{imp} \frac{\mu_n^2}{T^2} \frac {S (S+1)}{3} 
\sum_{{\bf r}_n} \frac {3 ( {\bf \mu}{_{imp}} {\bf r} ) ^2 +
 \mu_{imp}^2  r^2 }{r^8}.
\end{equation}

The sum  converges very rapidly. For fields less than
\begin{equation}
\label{9}
H_{00} = \frac{ \mu_{imp}}{ a^3} (n_{imp})^{1/2}
\end{equation}
 the main contribution to the specific heat comes from nuclei close to the
magnetic impurity. 

Consider next the range in $T$ and $H_0$ defined by the inequalities
\begin{eqnarray}
\label{10}
\mu H_0 \ll T \\
\frac{\mu_n \mu_{imp}}{a^3} \gg T \gg T_{cn}( = \frac {\mu_n^2}{a^3}).
\end{eqnarray}

In that region the main contribution to the specific heat comes from nuclei 
at large distance from the  impurity, i.e., $r \gg a$. For them one needs not accounting 
for the spin-spin interaction between nuclei (see  Eq. (\ref{10})) 
and one can also convert the summation over ${\bf r}{_n}$ into an integral of the form
 $ n_n \int d^3 {\bf r} ... \;$. From Eq. (\ref{5}) we obtain in this case 
\begin{equation}
\label{13}
\bar{C_n} - C_n^{(0)} = n_{imp} \frac{4\pi n_n \mu_{imp} \mu_n S}{3 T}  ( 1+ \frac{1}{2 \cdot 3^{1/2} } ln (2+ 3^{1/2}))
\end{equation}

By comparing Eqs. (\ref{6}) and (\ref{13}) one notices that for fields  
\begin{equation}
\label{14}
H_0 < (\frac { \mu_{imp} T n_{imp} }{\mu_n})^{1/2}
\end{equation}
the main contribution to the specific heat of the nuclear spin system comes from the interaction
with magnetic impurities. The  temperature dependence of this contribution is $T^{-1}$ rather then
$T^{-2}$. 
A dependence of this kind was indeed observed in Ref. \cite{herrmann} for $AuIn_2$.
We suggest that it is due to the impurity effect discussed here. However, for a quantitative comparison
one  must take into account that the main contribution to the specific heat comes  from
the $In$ nuclei which have spin $S=9/2$. The electric-field gradient due to the impurity leads
to a quadrupolar splitting of the spin levels. Being proportional to $r^{-3}$ the electric-field
gradient leads also to a $T^{-1}$ temperature dependence of the specific heat. For metals with nuclear
spin $S=1/2$ like $Ag$ [2] a quadrupolar splitting does, of course, not occur. 

A $T^{-1}$ contribution
results also from the nuclear spin-impurity spin RKKY-type of interaction $V_R$ 
\begin{eqnarray}
\label{15}
V_R = - \frac{ {\bf \mu}_n {\bf \mu_{imp}}}{r^3} \chi f(2 p_F r ) \; ;
f(x) = cos (x) - \frac {sin(x)}{x}.
\end{eqnarray}

Here $\chi$ is  a parameter proportional to the spin-spin
Fermi contact interaction \cite{statphys}. This interaction is
proportional  to the electronic charge 
at the  nucleus of the impurity \cite{quantmech}. 
For a light nucleus $\chi \ll 1$,
while for a heavy one  $\chi \gg 1$. 
Therefore in a metal the spin-spin interactions between the nucleus of an impurity
and the ones of the host contain two contributions given by Eqs. 
(\ref{1},\ref{15}). 
As pointed out before, we shall consider here only the universal dipole-dipole
interaction  (\ref{1}), which is the
same in  metals and insulators.

Next we consider the case of a strong magnetic field  $\mu_n H_0 \gg T$ . In this
regime the heat capacity is exponentially small  (see Eq. (\ref{3})). The
main contribution to it originates from nuclei for which the effective field
 $H({\bf r})$ is of order  $T / \mu_n$.
From Eq. (\ref{7}) we find  
\begin{equation}
\label{16}
H^2 = (\frac{ \mu_{imp}}{r^3} - H_0)^2 + \frac{3 \mu_{imp}}{r^3}
(\frac{\mu_{imp} }{r^3} +2 H_0) \frac {({\bf H}{_0} {\bf r})}{H_0 r}.
\end{equation}

One notices that $H({\bf r})$ is zero along a circle of radius  $r_0 = (\mu_{imp} /H_0)^{1/3}$
around the impurity in a plane perpendicular to ${\bf H}{_0} $. The nuclei contributing 
most to the specific heat are within a torus with its axis given by the circle of radius
$r_0$. If $\delta r$ denotes the distance from this axis and if  $z= ({\bf H}_0 {\bf r}) /H_0 r$
we find for the effective field
\begin{equation}
\label{17}
H^2 = 9 H_0^2 (z^2+ (\delta r /r_0)^2).
\end{equation}
With the help of Eqs. (\ref{3},\ref{17}) we obtain 

\begin{equation}
\label{18}
\bar{C_n}-C_n^{(0)} = \frac{ 16 \pi^2 T^2 \mu_{imp} n_n n_{imp}}{9 H_0^3 \mu_n^2}
(1- \frac{1}{(2 S +1)^2} ) I_1
\end{equation}
where
\begin{equation}
I_1 = \int_0^\infty \frac{dx \; x^3 }{sinh^2 x} = \frac{3}{2} \zeta (3), 
\end{equation}

and $\zeta(x)$ is Riemann's zeta function.

Replacing the sum over ${\bf r}{_n}$ by an integral $n_n \int d^3 {\bf r}$ is justified
only if many nuclei are placed within a radius   $\delta r /r_0 \simeq z \simeq T/ (\mu_n H_0)$
of the torus. 
This restriction leads to the requirement 
\begin{equation}
\label{19}
\frac{T^2 \mu_{imp} \mu_n n_n }{ (\mu_n H_0)^3} \gg 1 
\end{equation}

in order for Eq. (\ref{18}) to hold. Together with the starting assumption ( $\mu_n H_0 \gg T$)
this  implies the condition  
\begin{equation}
\label{20}
T \ll \mu_n H_0 \ll (T^2 \mu_{imp} \mu_n n_n)^{1/3}
\end{equation}
on the applied field $H_0$.

Next we deal with the case that the impurity spins are frozen in a glassy state.
Then Eqs. (\ref{3},\ref{5}) must be averaged over all directions of the external
field ${\bf H}{_0}$. As stated before, the main contribution to the nuclear heat capacity
comes from nuclei in an effective field  $H \simeq T/\mu_n$.
As a result we obtain 
\begin{equation}
\label{21}
\bar{C_n}-C_n^{(0)} =\frac{ 32 \pi T^3 \mu_{imp} n_n n_{imp}}{3 H_0^4 \mu_n^3}
(1+ \frac{1}{2{\sqrt 3}} ln(2+{\sqrt 3}))(1-\frac{1}{(2S+1)^3})I_2,
\end{equation}
where 
\begin{equation}
I_2 = \int_0^\infty \frac{dx \; x^4}{sinh^2 x} = \frac{\pi^4}{30}.
\end{equation}

The restriction  (\ref{20}) for $H_0$ is changed accordingly into 
\begin{equation}
\label{22}
T \ll  \mu_n H_0 \ll (T^3 \mu_{imp} \mu_n n_n )^{1/4}.
\end{equation}

The required range of strong magnetic fields has not yet been studied experimentally,
although in Ref. \cite{herrmann} the region $\mu_n H_0 \leq T$ was investigated.
For metals in the clean limit the nuclear-spin contribution to the heat capacity
has a maximum near  $\mu_n H_0 \sim T$,  the position of which is only weakly dependent
on the magnetic impurity concentration $n_{imp}$, provided  $n_{imp} \mu_{imp} / \mu_n \ll n_n$.
We expect that in the strong-field regime $\mu_n H_0 \gg T$  the heat capacity has a power-law
behaviour given by Eqs. (\ref{18}) and (\ref{21}). For pure samples an exponential temperature
dependence is obtained  (see Eq. (\ref{3})).

In summary, we have shown that in a wide region of temperature and magnetic fields the main 
contribution to the nuclear specific heat results from their interaction with small amounts
of magnetic impurities which are present in most of the systems.

{\bf Acknowledgements}

Two of us (Yu. N. O. and A. M. D.) wish to acknowledge die Max-Planck-Gesellschaft 
zur F\"orderung der Wissenschaften and Max-Planck-Institut f\"ur Physik komplexer
Systeme for the hospitality during the period of doing  this work.

The research  of Yu. N. Ovchinnikov was supported by the CRDF grant RP1-194 and 
by the Naval Research Lab contract N 00173-97-P-3488.

The research of A.M. Dyugaev was supported by the INTAS-RFBR 95-553 grant.

\end{document}